\documentclass[11pt,twocolumn]{article} % for two column format

\usepackage{graphics,graphicx}

\newcommand{\xmm} {{XMM-Newton}}
\newcommand{\chandra} {{Chandra}}

\newcommand{\cmsq} {cm$^{-2}$}
\newcommand{\nh} {$N_{\rm{H}}$}
\newcommand{\lx} {$L_{\rm{X}}$}
\newcommand{\fx} {$F_{\rm{X}}$}
\newcommand{\chisq} {$\chi^2$}
\newcommand{\dchisq} {$\Delta\chi^2$}

\newcommand{\mic}{{${\mu}$m}}

\newcommand{\degree}{{$^\circ$}}

\newcommand{\ergs}{\mbox{\thinspace erg\thinspace s$^{-1}$}}
\newcommand{\ergcms}{\mbox{\thinspace erg\thinspace cm$^{-2}$\thinspace s$^{-1}$}}

\newcommand{\msol} {$M_{\odot}$}

\title{Magnetic field strength of a neutron-star-powered ultraluminous X-ray source}

\author{M. Brightman$^{1*}$, F. A. Harrison$^{1}$, F. F\"{u}rst$^{2}$, M. J. Middleton$^{3}$, D. J. Walton$^{4}$, D. Stern$^{5}$,\\ A. C. Fabian$^{4}$, M. Heida$^{1}$, D. Barret$^{6,7}$, M. Bachetti$^{8}$}

\begin{document}

\maketitle

{\footnotesize
$^{1}$Cahill Center for Astrophysics, California Institute of Technology, 1216 East California Boulevard, Pasadena, CA 91125, USA\\
$^{2}$European Space Astronomy Centre (ESAC), Science Operations Departement, 28692 Villanueva de la Ca\~{n}ada, Madrid, Spain\\
$^{3}$Department of Physics and Astronomy, University of Southampton, Highfield, Southampton SO17 1BJ, UK\\
$^{4}$Institute of Astronomy, University of Cambridge, Madingley Road, Cambridge CB3 0HA, UK\\
$^{5}$Jet Propulsion Laboratory, California Institute of Technology, 4800 Oak Grove Drive, Pasadena, CA 91109, USA\\
$^{6}$CNRS, IRAP, 9 avenue du Colonel Roche, BP 44346, F-31028 Toulouse Cedex 4, France\\
$^{7}$Universit\'{e} de Toulouse, UPS-OMP, IRAP, Toulouse, France\\
$^{8}$INAF, Osservatorio Astronomico di Cagliari, Via della Scienza 5, 09047 Selargius, Italy\\}

{\bf 
Ultraluminous X-ray sources (ULXs) are bright X-ray sources in nearby galaxies not associated with the central supermassive black hole. Their luminosities imply they are powered by either an extreme accretion rate onto a compact stellar remnant, or an intermediate mass ($\sim100-10^5$~\msol) black hole\cite{farrell09}. The recent detection of coherent pulsations coming from three bright ULXs\cite{bachetti14,fuerst16,israel17,israel17a} demonstrates that some of these sources are powered by accretion onto a neutron star, implying accretion rates significantly in excess of the Eddington limit, a high degree of geometric beaming, or both. The physical challenges associated with the high implied accretion rates can be mitigated if the neutron star surface field is high - in the magnetar regime ($10^{14}$~G)\cite{dallosso15}, since this suppresses the electron scattering cross section, reducing the radiation pressure that chokes off accretion for high luminosities. One of the few ways to determine surface magnetic fields is through the detection of cyclotron resonance scattering features (CRSFs)\cite{gnedin74,truemper78} produced by the transition of charged particles between quantized Landau levels. To date, CRSFs have only been detected in Galactic accreting pulsars. Here we present the detection  at 3.8-$\sigma$ significance of a strong absorption line at a rest-frame energy of 4.5~keV in the Chandra spectrum of a ULX in M51. We find that this feature is likely to be a CRSF produced by the strong magnetic field of a neutron star. Assuming scattering off electrons, the magnetic field strength is implied to be $\sim10^{11}$~G, however the line is narrower than any electron CRSFs previously observed, and assuming thermal broadening, the implied temperature is significantly cooler than the accretion disk or column. The line shape is, however, consistent with a proton resonance scattering feature, implying that the neutron star has a magnetic field near the surface of B$\sim10^{15}$~G.
}

The interacting galaxies of M51 have a large population of ULXs which were first discovered by the Einstein X-ray Observatory\cite{palumbo85}. Since the galaxies are nearby (8.58 Mpc)\cite{mcquinn16} and face on, their ULXs have been well studied and characterised by many X-ray telescopes including the X-ray Multi-Mirror Mission (XMM) Newton and Chandra observatories. RX J133007+47110, otherwise known as NGC5194/5 ULX8 [ref.\cite{liu05}] is located at 13h 30m 07.55s, +47\degree\ 11' 06.1'' on the sky (J2000 coordinates) and is $\sim$2.6~arcmin east of the nucleus of M51a (NGC 5194) in one of the spiral arms (Figure 1a). It is one of the brightest ULXs in M51 with an X-ray luminosity of $\sim2\times10^{39}$~\ergs\ (assuming isotropic emission)\cite{dewangan05}.

On 2012 September 09 the Chandra X-ray observatory conducted a deep 181~ks observation of M51 with its Advanced CCD Imaging Spectrometer (ACIS) during which ULX8 was brighter than previously observed and exhibited several $\sim10$-ks duration bursts of X-rays up to $10^{40}$~\ergs\ that occurred throughout the observation (Figure 1b). We extracted and fitted the ACIS spectrum of ULX8 from the entire observation, which revealed a strong absorption line feature at rest-frame 4.52$\pm$0.05~keV, with a Gaussian width, $\sigma=0.11^{+0.07}_{-0.05}$~keV, and an equivalent width, EW$=-0.19^{+0.06}_{-0.09}$~keV. The continuum is well described by a power-law spectrum ($\Gamma=1.3\pm0.3$) with an exponential cut off ($E_{\rm C}=3.7^{+2.2}_{-1.0}$~keV, Figure 1c) or a multi-colour disk model with temperature $T=1.87^{+0.54}_{-0.28}$~keV and radial temperature profile with index $p=0.54\pm0.03$, typical of other ULXs observed below 10~keV [ref.\cite{gladstone09}]. The goodness of fit for these models are \chisq=252.4 and 253.2 respectively, both with 204 degrees of freedom. The addition of a Gaussian absorption line improves the cut-off power-law fit by \dchisq$=-44.3$, which from simulations we find is highly significant ($\sim3.8\sigma$). 

The absorption line is also present throughout the observation and does not appear to change during the bursting episodes. We also searched through all other long archival XMM-Newton and Chandra observations of M51 to determine if the absorption line is transient, or a permanent feature of the source. We modeled the spectra in the same way as described above but with the line width fixed to 0.1~keV and the energy allowed to vary by $\pm1$~keV from 4.5~keV. We do not find that the addition of the Gaussian absorption line is significant in any other observation, but the statistical lower limits on the EW afforded by the lower signal to noise of these observations are such that we cannot rule out its presence.

\begin{figure*}
\begin{center}
\includegraphics[width=140mm]{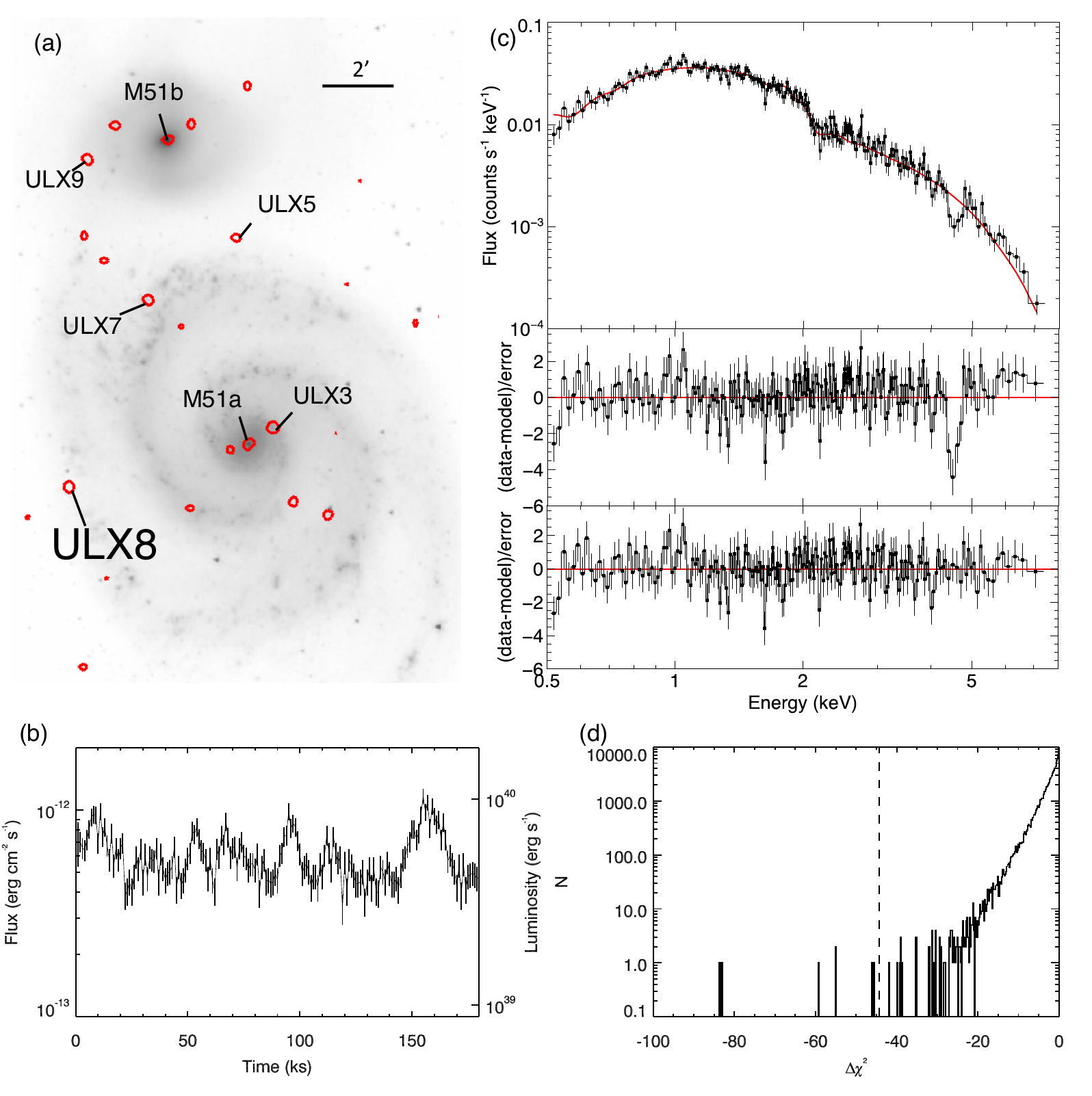}
\caption{(a) The position of ULX8 in M51. The grey-scale image is a near-infrared (Spitzer/IRAC 3.6\mic) image of the galaxy and the red circles are Chandra contours showing the brightest X-ray sources ($>2$ counts pixel$^{-1}$) . The nuclei of the two galaxies and the ULXs are labelled. (b) is the 0.5--8~keV Chandra lightcurve from observation 13813 with time bins of 1000s and 1$\sigma$ error bars. (c) The top panel is the Chandra spectrum (black data points, 1$\sigma$ error bars) from the same observation fitted with an absorbed cut-off power-law model (red line), the middle panel shows the residuals to the fit showing clearly the absorption line at 4.5~keV and the bottom panel shows the residuals after the addition of a Gaussian absorption line. The horizontal red line again represents the fitted model (d) is the results from the significance simulations showing a histogram of \dchisq\ values obtained from adding a line to 10000 simulated spectra. The dashed line shows the observed change in \chisq.}
\label{figure1}
\end{center}
\vspace{-10mm}
\end{figure*}

There are no absorption features related to the Chandra mirrors or detectors around 4.5~keV and the count rate is low enough that pileup effects that could potentially distort the spectrum are small (the pileup fraction is $<10$\%). Therefore we rule out an instrumental origin of the absorption line. While several atomic absorption and emission lines from iron, oxygen and neon have recently been detected from stacked XMM-Newton and NuSTAR observations of three other ULXs\cite{pinto16,walton16}, the energy of the line we find in the Chandra spectrum of ULX8 is not consistent with any known bound-bound atomic transition. Highly blue shifted features from lower energies caused by an ultra-fast outflow are unlikely since the line observed is isolated whereas the low energy lines are usually accompanied by lines from other elements. The large EW of the line makes absorption by iron, with its high cosmological abundance, the most likely atomic solution. However, the nearest strong iron absorption line is the Fe {\sc xxv} K$\alpha$ transition at 6.67~keV, an association with which would require a redshift of $\sim$0.5. At the flux of ULX8, less than 0.15 galactic and extragalactic sources are expected in the region of M51 [ref.\cite{palumbo85}]. Given this expected rate, the probability that ULX8 is a background AGN or a foreground source based on a Poisson distribution is $<0.13$. Furthermore, the near infrared (NIR) fluxes of AGN are as high or higher than their X-ray fluxes\cite{lusso11}, even considering a redshift of 0.5, but the upper limit on the NIR emission at the position of the ULX is a magnitude lower than the observed X-ray flux\cite{heida14}. 

\begin{table*}
\centering
\label{tab_obsdat}
\begin{center}
\begin{tabular}{l l c r}
\hline
Observatory	& ObsID	& Start date (UT)	& Exposure \\
			&		&				& (ks)	   \\
\hline

\xmm\ &0112840201&2003-01-15 13:12:55& 20.9\\
\xmm\ &0212480801&2005-07-01 06:38:00& 49.2\\
\xmm\ &0303420101&2006-05-20 06:31:01& 54.1\\
\xmm\ &0303420201&2006-05-24 11:12:05& 36.8\\
\chandra\ &13813&2012-09-09 17:47:30&181.6\\
\chandra\ &13812&2012-09-12 18:23:50&159.5\\
\chandra\ &13814&2012-09-20 07:21:42&192.4\\
\chandra\ &13815&2012-09-23 08:12:08& 68.1\\
\chandra\ &13816&2012-09-26 05:11:40& 74.1\\

\hline
\end{tabular}
\caption{Details of the observations used in this analysis.}

\end{center}
\end{table*}

The best remaining possibility is that this line is produced by a CRSF, since these lines can theoretically occur over a wide range of energies depending mostly on the strength of the magnetic field which causes them. Most CRSF features which have been observed in Galactic neutron star binaries are associated with transitions of electrons, such as SMC X-2 [ref.\cite{jaisawal16}] and V0332+53 [ref.\cite{tsygankov06}]. Alternatively, some CRSF features have been ascribed to the transitions of protons, most notably in the spectra of soft $\gamma$-ray repeaters such as SGR~1806-20 [ref.\cite{ibrahim02}] and SGR~0418+5729 [ref.\cite{tiengo13}]. SGRs are sources which emit large bursts of X-rays and $\gamma$-rays and were shown to be powered by magnetars\cite{kouveliotou98}. Association with either electrons or protons gives significantly different measurements of the magnetic field strength, however. For electrons, the transition energy, $\Delta E$, is $11.6(1+z)^{-1}(B/10^{12}$ G)~keV, where $z=(1-2GM/r_{cyc}c^2)^{-1/2}-1$ is the gravitational redshift and is $\sim0.25$ for the emission from the surface of a typical neutron star. The 4.5~keV line that we have detected would therefore imply $B=4(1+z)\times10^{11}$ G. For protons, $\Delta E=6.3(1+z)^{-1}(B/10^{15}$ G)~keV, thus, interpreting our observed line as a proton CRSFs would imply a very high magnetic field strength of $7(1+z)\times10^{14}$ G.

\begin{table*}
\centering
\label{tab_specpar}
\begin{center}
\begin{tabular}{c c c c c c c}
\hline
obsID & \nh & $\Gamma$ & $E_{\rm C}$ &\chisq/DoFs & \fx\ & \lx \\
 & ($10^{20}$ \cmsq) & & (keV) & & ($10^{-13}$ \ergcms) & ($10^{39}$ \ergs) \\
(1) & (2) & (3) & (4) & (5) & (6) & (7)  \\
\hline
   112840201& 11.2$^{+  3.3}_{-  4.5}$ & 2.27$^{+ 0.08}_{- 0.44}$ &- & 72.1/58& 2.99$^{+ 0.28}_{- 0.52}$ &  2.6$^{+  0.2}_{-  0.5}$ \\
   212480801&  6.0$^{+  2.8}_{-  2.5}$ & 1.99$^{+ 0.15}_{- 0.19}$ &- & 82.7/66& 2.49$^{+ 0.16}_{- 0.16}$ &  2.2$^{+  0.1}_{-  0.1}$ \\
   303420101&  9.9$^{+  2.8}_{-  2.6}$ & 2.19$^{+ 0.16}_{- 0.21}$ &- & 78.1/79& 2.20$^{+ 0.14}_{- 0.14}$ &  1.9$^{+  0.1}_{-  0.1}$ \\
   303420201& 13.2$^{+  3.3}_{-  3.4}$ & 2.25$^{+ 0.15}_{- 0.28}$ &- & 65.8/62& 2.69$^{+ 0.21}_{- 0.22}$ &  2.4$^{+  0.2}_{-  0.2}$ \\
       13812&  4.7$^{+  4.8}_{-  4.7}$ & 1.12$^{+ 0.40}_{- 0.40}$ &  2.1$^{+  0.9}_{-  0.5}$ &170.2/153& 3.30$^{+ 0.31}_{- 0.26}$ &  2.9$^{+  0.3}_{-  0.2}$ \\
       13813&  7.7$^{+  3.7}_{-  3.7}$ & 1.26$^{+ 0.29}_{- 0.30}$ &  3.7$^{+  2.2}_{-  1.0}$ &252.4/204& 5.43$^{+ 0.35}_{- 0.32}$ &  4.8$^{+  0.3}_{-  0.3}$ \\
       13814& 10.5$^{+  5.4}_{-  5.3}$ & 1.70$^{+ 0.46}_{- 0.47}$ &  2.5$^{+  2.0}_{-  0.8}$ &145.6/141& 2.64$^{+ 0.37}_{- 0.30}$ &  2.3$^{+  0.3}_{-  0.3}$ \\
       13815& 16.7$^{+  8.4}_{-  9.7}$ & 2.30$^{+ 0.52}_{- 0.92}$ &  $>1.8$ & 69.6/73& 3.13$^{+ 0.65}_{- 0.70}$ &  2.8$^{+  0.6}_{-  0.6}$ \\
       13816&  4.4$^{+  8.1}_{-  4.4}$ & 1.12$^{+ 0.71}_{- 0.50}$ &  2.1$^{+  2.4}_{-  0.8}$ &112.2/99& 2.98$^{+ 0.54}_{- 0.32}$ &  2.6$^{+  0.5}_{-  0.3}$ \\
\hline
\end{tabular}
\caption{{ Parameters of the cut-off power-law model used to fit the continuum, without the absorption line.} Column (1) list the observation identification number, column (2) lists the neutral column density intrinsic to the source, column (3) lists the power-law index of the continuum model and column (4) lists the high energy cut off. A `-' indicates the parameter is unconstrained. Column (5) lists the \chisq\ of the fit and the number of degrees of freedom, column (6) lists the 0.5--8~keV flux, corrected for absorption and column (7) lists the intrinsic luminosity of the source assuming isotropic emission and a distance of 8.58 Mpc.}
\end{center}
\end{table*}

One piece of observational evidence that could serve to discriminate between these two scenarios is the width of the absorption feature, which is thought to be caused by the Doppler effect produced by the thermal energy of the charged particles. Electrons, which are much lighter than protons, produce broader lines. For Galactic pulsars with known electron CRSFs, the features are typically broad with Gaussian line widths of order $\sim1$~keV, and are seen mostly at energies above 10~keV, giving broadening ratios, $\sigma/E\sim0.1$\cite{tsygankov06,jaisawal16}. Protons on the other hand are more massive and should produce narrower lines. The few proton CRSF observed to date were indeed narrow ($\sigma<0.4$~keV) and at energies below 10~keV\cite{ibrahim02}, giving broadening ratios of $\sigma/E<0.1$. The broadening ratio of the line we have observed is 0.02 and therefore more comparable to the previously reported proton CRSFs. However, ULX8 is the most luminous CRSF source identified so far, so drawing a connection to lower luminosity neutron star systems may be tenuous.

The velocity dispersion, $v$, associated with the line width we have observed is $0.02c=6\times10^{6}$ m\,s$^{-1}$ ($\frac{\sigma}{E}=\frac{v}{c}$). For electrons, the implied thermal energy ($\frac{1}{2}m_{e}v^2$) is $1.6\times10^{-17}$ J (0.1~keV), whereas the implied thermal energy when considering protons is $\sim200$~keV. The temperature of the accretion disk, which we have measured from the continuum, is 2~keV and is not consistent with either scenario. However, it is not known where the CRSF is formed in this system, and may be a hotter region of the system, or indeed a cooler one.

An additional key piece of evidence for CRSFs is the detection of harmonic features caused by transitions to higher Landau levels and are expected at integer multiples of the energy of the fundamental line. Assuming the 4.5~keV line is the fundamental, the first harmonic should lie at 9.0~keV. However, the effective area of the Chandra detectors is very low at this energy ($\sim10$\% of the area at 4.5~keV) so no constraints can be placed on the presence of a harmonic line here. Nevertheless, we can investigate if the 4.5~keV line is in fact a harmonic, by testing the presence of a fundamental line at 2.25~keV. Adding an absorption line at this energy with the same width as the 4.5~keV line does not improve the fit (\dchisq$<0.1$). If it were present, it would have an equivalent width of $-5\times10^{-5}$~keV (90\% confidence lower limit of -0.03~keV) and thus several factors smaller than the line at 4.5~keV. If there is indeed a fundamental line at 2.25~keV, it has suffered from a large amount of photon spawning, caused by transitions from high to low Landau levels that produce photons at the energy of the fundamental line\cite{araya99}. This effect is strongest for electrons and for hard X-ray spectra, whereas the spectrum of ULX8 is relatively soft ($\Gamma=1.3\pm0.3$ with an exponential cut off at 4~keV). We therefore conclude that the 4.5~keV line is the likely fundamental. The detection of a harmonic line at 9~keV and measuring its strength with respect to the fundamental could provide the key observational evidence in favour of an electron or proton CRSF. This could be done with NuSTAR, \xmm\ or NICER when the ULX exhibits a similar bursting bursting behavior as it did when the CRSF was detected by \chandra.

While the CRSF detected unambiguously determines that the compact object that powers ULX8 is a neutron star we do not detect pulsations from this source. If the neutron star is spinning, and the rotation period is 1\,s, we find that the pulse fraction would need to be at least 45\% in order to be detected at 3-$\sigma$ significance in the short \xmm\ observations. This is higher than the pulsed fractions below 10~keV of the other ULX pulsars discovered to date, which are $<30$\%\cite{bachetti14,fuerst16,israel17,israel17a}. Although the \chandra\ observations are longer, the time resolution of the ACIS detectors is not good enough to detect the $\sim1$-s pulsation periods typical of the other ULX pulsars.

To date the only estimates of the magnetic field strengths of neutron-star-powered ULXs have come from the observed spin up of the neutron star caused by the transfer of angular momentum from the accretion disk to the star. The accretion torque is applied at the inner edge of the accretion disk which is located where the magnetic field disrupts the accretion flow. This radius depends on the magnetic field strength, so knowledge of the spin up combined with the accretion rate gives a magnetic field strength estimate. These estimates have generally yielded strengths of B$\sim10^{12}$ G for the known ULX pulsars\cite{bachetti14,fuerst16} and are thus consistent with the magnetic field strength inferred for ULX8 when assuming an electron CRSF. Other theoretical arguments have hypothesized strengths that vary over five orders of magnitude, from $10^9-10^{14}$ G. The proton scenario yields a magnetic field strength of B$\sim10^{15}$ G and supports models that invoke such strong magnetic fields that are needed to reduce the electron scattering cross section and allow super-critical accretion\cite{dallosso15}. In order to reconcile this measurement with the magnetic field implied by the spin up, the strong field which produces the CRSF may be multipolar, while the dipolar outer field, which controls the spin-up behaviour, could be much weaker\cite{israel17,israel17a}.

\section{Methods}

\subsection{Observations and data reduction.}
{\bf Chandra} has observed M51 on 13 separate occasions between 2000 and 2017. The majority of the exposure time resulted from a large 750-ks program to study the interstellar medium, supernova remnants and X-ray binaries in the galaxy$^{\rm 28}$. We consider only five of all 13 observations where the exposure is longer than 50 ks such that the signal to noise is large enough for detailed spectral fitting. We use {\sc ciao} v4.7 to analyze these data. We extract events from a circular region with a 1 arcsec radius centered on the source, and background events from a nearby source-free region using the tool {\sc specextract}.

{\bf XMM-Newton} has observed M51 nine times between 2003 and 2011. We analyze only four of these which have exposures longer than 20 ks using {\sc xmmsas} v15.0 software. Events are extracted from a 15 arcsec circular region and background from a larger region located on the same chip. We only consider EPIC-pn data due to its larger effective area than the EPIC-MOS instruments and filter out periods of high background from the \xmm\ data. We list the observational data in Table 1.

\subsection{Spectral fitting.}
We grouped all spectra with a minimum of 20 counts per bin with the {\sc heasoft} tool {\sc grppha}. Spectral fitting was carried out using {\sc xspec} v12.9.1 to background-subtracted spectra using \chisq\ as the fit statistic. For the \chandra\ spectra we fit in the 0.5--8~keV range and for XMM-Newton we fit in the 0.2--10~keV range. 

We fit the \chandra\ and \xmm\ data with a power-law model with a high-energy cut off, where $F_{\gamma}=AE^{-\Gamma}e^{(-E/E_{\rm C})}$, a phenomenological model that reproduces the spectral curvature typical of ULXs below 10~keV. We also apply photoelectric absorption caused by the interstellar medium of our own Galaxy, fixed at \nh=$1.5\times10^{20}$ \cmsq, as well as absorption intrinsic to the source at the redshift of M51 ($z=$0.002), left as a free parameter in the fit.

Table 2 lists the results from the spectral fitting. We find that the absorbed cut-off power-law model describes the data well in general. The parameters yielded vary from observation to observation, with \nh$=4-16\times10^{20}$ \cmsq\ and power-law index $\Gamma=1.1-2.3$. $E_{C}$ is not well constrained in many observations, but when it is, it varies from $E_{\rm C}=2-5$~keV. The absorption corrected luminosity ranges from $2-5\times10^{39}$ \ergs. Inspecting the residuals of the fits, we discovered the strong absorption-line feature at 4.5~keV in the Chandra observation 13813. 

\subsection{Significance simulations.}
In order to determine the statistical significance of the absorption line, we carry out simulations. We take the best fit absorbed cut-off power-law model from the Chandra observation and simulate $10^5$ spectra in {\sc xspec} using the {\tt fakeit} command. We conduct these simulations based on the RMF and ARF files for this observation. We simulate source and background spectra with the same exposure time as the real observation. 

We then refit the spectrum with the same continuum model and note the \chisq\ of the fit. We then add the Gaussian absorption line with an initial trial energy of 4.5~keV and the Gaussian width fixed to 0.1~keV. The equivalent width was allowed to take any negative value. The energy of the line was allowed to vary between 0.5 and 8~keV. We then note the \dchisq\ of the fit after the addition of this line. A histogram of the \dchisq\ values produced by the simulated data is presented in Figure 1d, where the real \dchisq\ is plotted as a dashed line. From the $10^5$ simulated spectra, 18 produced a \dchisq\ as large as the real observation. Varying the initial energy of the line in the fit does not change the result. The false alarm rate of 2$\times10^{-4}$ corresponds to a significance of $\sim3.8\sigma$.

\subsection{Absorption line characteristics.}
While the absorption line feature at 4.5~keV is not visually evident in the other data sets, we add the absorption line to these spectral fits in order to determine a lower limit to the EW to assess if the line is a transient feature. As for the significance simulations, we fix the width of the line to 0.1~keV. We allow the energy of the line to vary by $\pm1$~keV from 4.5~keV to allow for energy dependence e.g. on luminosity\cite{tsygankov06}. We find only two out of nine observations where the 68\% confidence intervals on the EW of the line is not consistent with the EW width of the strongly detected line ($-0.18^{+ 0.04}_{- 0.09}$~keV). If we stack the \chandra\ observations where the line was not detected and the ULX was at low fluxes, we can place a tighter constraint on the lower limit of the EW of $-0.1$~keV at 90\% confidence. Doing the same with the \xmm\ observations yields a lower limit of -0.2~keV. This indicates that the line is not necessarily transient but may be weaker at lower fluxes and require high signal to noise spectra in order to detect. 

During Chandra observation 13813 the flux from ULX8 is highly variable, ranging from 5$\times10^{-13}$ \ergcms\ to $10^{-12}$ \ergcms. We test if the properties of the absorption line change during the flaring episodes by splitting the observation into two flux regimes, a low flux regime where \fx$<6\times10^{-13}$ \ergcms\ and a high flux regime where \fx$>6\times10^{-13}$ \ergcms. We find the continuum parameters in the low flux regime are $\Gamma=1.31^{+0.34}_{-0.35}$ and $E_{\rm C}=3.2^{+2.2}_{-1.0}$~keV, and the line parameters are E=4.54$^{+0.07}_{-0.17}$~keV, $\sigma<0.48$~keV and EW=$-0.16^{+0.08}_{-0.34}$~keV. During the high flux state, the continuum parameters are $\Gamma=1.46^{+0.35}_{-0.37}$ and $E_{\rm C}=4.5^{+53}_{-3.6}$~keV and the line parameters are E=4.46$^{+0.22}_{-0.13}$~keV, $\sigma=0.21^{+0.48}_{-0.13}$~keV and EW=$-0.21^{+0.11}_{-0.35}$~keV. We find no evidence for any spectral changes in either the line or continuum during the flaring episodes.

\subsection{Timing analysis.}
For each of the observations where we conducted line searches, we also searched for pulsations expected from a neutron star accretor. We applied barycentric corrections to each events file, using the {\sc ciao} tool {\sc axbary} on the \chandra\ data and the {\sc xmmsas} tool {\sc barycen} on the \xmm\ data.  Unbinned source+background lightcurves for the ULX were constructed and fast -Fourier transform (FFT) analysis was carried out on the lightcurves using the {\sc idl} tool {\sc fft}. However, no evidence for coherent pulsations was found.

Finally we proceed to determine the upper limit on the putative pulsed fraction by simulating a light curves with the same statistical properties as the measured ones, but with an additional sinusoid with a 1-s period that has some pulsed fraction on top of it. From this theoretical light curve we draw Poisson-distributed measurements and calculate a FFT on each of the randomized lightcurves. We then determine if any power in these FFTs is higher than the 3$\sigma$ limit for white noise, taking the number of trials (i.e., frequencies in the power spectrum density) into account. We run this 500 times for four pulsed fractions, 30, 35, 40 and 45\% finding that 20.4, 62, 92 and 99.8\% of these lightcurves show pulsations at $>3\sigma$ respectively. Therefore this ULX would require a high pulsed fraction in order for the pulsations to be detected at high confidence with the current observations.

\subsection{Acknowledgements} 
M.J.M. and D.J.W. appreciate support from Ernest Rutherford STFC fellowships. The work of D.S. was carried out at the Jet Propulsion Laboratory, California Institute of Technology, under a contract with NASA.

\bibliography{bibdesk.bib}

\end{document}